\documentclass[aps,prc,twocolumn,superscriptaddress,showpacs,floatfix,nofootinbib]{revtex4-1}
\usepackage{amsmath,graphicx}

\begin{document}
\preprint{SPhT-t10/?}

\title{Directed flow at midrapidity in heavy-ion collisions}

\author{Matthew Luzum}
\affiliation{
CEA, IPhT, Institut de physique th\'eorique de Saclay, F-91191
Gif-sur-Yvette, France} 
\author{Jean-Yves Ollitrault}
\affiliation{
CNRS, URA2306, IPhT, Institut de physique th\'eorique de Saclay, F-91191
Gif-sur-Yvette, France} 
\date{\today}

\begin{abstract}
It was recently shown 
that fluctuations in the initial
geometry of a heavy ion collision 
generally result in
a dipole asymmetry of the distribution of outgoing particles. 
This asymmetry, unlike the usual directed flow, is expected to be present at a wide range of rapidity --- including midrapidity. First evidence of this phenomenon can be seen in recent two-particle
correlation data by STAR, providing the last element necessary to quantitatively describe long-range dihadron correlations.  We extract differential directed flow from these data and propose a new direct measurement.
\end{abstract}


\maketitle

\section{Introduction}
Analyses of correlations between particles emitted in heavy-ion collisions reveal 
azimuthal structure that is not seen in proton-proton or deuteron-gold 
collisions. These unique correlations are present even if the particles are separated by a 
large interval in rapidity. The largest component, known as elliptic flow, is one 
of the early observations at RHIC~\cite{Ackermann:2000tr}. 
More recently, it was noticed that more detailed features of the correlation
pattern known as ``ridge'' and ``shoulder'' can also be explained by a new phenomenon, 
dubbed ``triangular flow'',  which comes from a hydrodynamic response to  fluctuations 
in the initial geometry~\cite{Alver:2010gr}. 
An analysis~\cite{Luzum:2010sp} of recent experimental data~\cite{Agakishiev:2010ur}
shows that {\it all\/} 
the correlations observed at large relative pseudorapidities are likely to 
originate from collective flow (and global momentum conservation).  Contributions from event-by-event fluctuations, which in the past have been neglected, are key to this understanding.  

Teaney and Yan~\cite{Teaney:2010vd} have shown that such fluctuations in the initial
geometry are expected to create a new type of directed flow in addition to elliptic and triangular flow. 
Fluctuations break the symmetry of the initial density profile, and as a result there 
is, in general, one direction where the profile is steepest. 
This effect can be quantified as a dipole asymmetry in the initial density~\cite{Teaney:2010vd}:
\begin{equation}
\label{defpsi1}
\varepsilon_1 e^{i\Phi_1}=-\frac{\langle r^3 e^{i\phi}\rangle}{\langle r^3\rangle}.
\end{equation}
where the averages in the right-hand side are taken over the initial transverse entropy density profile, and 
$(r,\phi)$ is a polar coordinate system around the center of the distribution, chosen such that 
$\langle r e^{i\phi}\rangle=0$.
If one chooses $\varepsilon_1$ to be positive, then $\Phi_1$ generally corresponds 
to the steepest direction for a smooth profile, and $\varepsilon_1$ is the 
magnitude of the dipole asymmetry. In general,  $\varepsilon_1$ will differ from 0
 --- even at midrapidity --- due to fluctuations. 

This dipole asymmetry, followed by hydrodynamic expansion, creates 
a specific type of flow.  Recall that particles with large transverse momentum $p_t$ are 
created where the fluid velocity is largest, and their momentum is parallel to the fluid 
velocity~\cite{Borghini:2005kd}. The fluid velocity scales like the gradients of the 
initial density profile, and so high-$p_t$ particles are likely to be emitted along the 
steepest gradient, i.e., with azimuth $\Phi_1$.
Just as the quadrupole asymmetry in the distribution of outgoing particles is quantified by the well-studied elliptic flow coefficient $v_2$,  the second Fourier harmonic with respect to the event plane, this asymmetry can be characterized by the directed flow measured with respect to
$\Phi_1$
\begin{equation}
\label{defv1}
v_1\equiv\langle\cos(\phi_p-\Phi_1)\rangle,
\end{equation}
where $\phi_p$ is the azimuthal angle of the outgoing particle momentum and the brackets indicate an average over these particles in a given collision event.

While $v_1$ is positive for high-$p_t$ particles, the
condition that the total net transverse momentum of the fluid approximately vanishes implies 
in turn that $v_1$ is negative for low-$p_t$ particles, which results in a specific
pattern for the $p_t$ dependence of $v_1$~\cite{Teaney:2010vd}.
Although $\Phi_1$ cannot be measured experimentally, 
the correlation of every particle to $\Phi_1$ induces a correlation
between pairs of particles, which can be measured even if $\Phi_1$ is
not known. 

Unlike the usual directed flow, which is odd in rapidity~\cite{Alt:2003ab,Adams:2003zg,Back:2005pc},
this new type of directed flow is expected to depend little on rapidity, since it is created by 
fluctuations in the initial geometry. 
Hints of this phenomenon can already be seen in 
existing data, and it enables a complete quantitative understanding of long-range dihadron correlations (in conjunction with the previously studied elliptic flow, quadrangular flow \cite{Adams:2003zg}, triangular flow, and momentum conservation) \cite{Luzum:2010sp}.
In this article, we extract differential directed flow from recent experimental data~\cite{Agakishiev:2010ur}, and we propose a method to measure it directly.

\section{Dihadron correlations}
\label{s:star}

The STAR collaboration has released an analysis of dihadron
correlations at different angles with respect to the event
plane~\cite{Agakishiev:2010ur}.
They have measured the distribution of the relative azimuthal angle 
$\Delta\phi$
between a high-$p_t$ trigger particle and an associated 
particle in various $p_t$ bins. 
 
Along with the contribution from flow, this measurement also contains a correlation induced by the global conservation of transverse momentum.  The overall probability of detecting a pair of particles with transverse momenta ${\bf p}_t^{(t)}$ and ${\bf p}_t^{(a)}$  
can be written in terms of the one-particle distribution $dN/d^2{\bf p}_t$, under fairly general assumptions, as~\cite{Borghini:2000cm}:
\begin{equation}
\label{eq:dndp}
\frac {dN_{pairs}} {d^2{\bf p}_t^{(t)}d^2{\bf p}_t^{(a)}} = \frac {dN} {d^2{\bf p}_t^{(t)}} \frac {dN} {d^2{\bf p}_t^{(a)}} \left( 1 - 2 \frac{{\bf p}_t^{(t)}\cdot {\bf p}_t^{(a)}} {\left\langle  \sum\ p_t^2 \right\rangle} \right).
\end{equation}
where the sum is over all particles in the event and the brackets average over all events in the centrality bin under consideration, and $p_t = \sqrt{{\bf p}_t^2}$.
The term in parentheses represents the preference for back-to-back emission due to the near-vanishing total transverse momentum.

The one-particle distribution is typically characterized in terms of Fourier components
\begin{equation}
\frac {dN} {d^2{\bf p}_t} = \frac N {2\pi} \left( 1 + 2 \sum v_n \cos n(\phi_p - \psi_n) \right).
\end{equation}
Plugging this into Eq.~\eqref{eq:dndp}, the first Fourier harmonic of this two-particle correlation,
$\langle\cos\Delta\phi\rangle$,
can then be written, to first order, as a sum of contributions from momentum conservation and from directed flow~\cite{Luzum:2010sp}
\begin{equation}
\label{decomposition}
\left\langle\cos\Delta\phi\right\rangle=v_{1}^{(t)} v_{1}^{(a)} + \left\langle\cos\Delta\phi\right\rangle_{\rm p_t\ cons.}.
\end{equation}
Here, $v_{1}^{(t)}$ and $v_{1}^{(a)}$ are the directed flows of the associated and trigger particles, and higher order terms are negligible when each of these is much smaller than 1.  Explicitly, then, the momentum conservation correlation is
  \begin{equation}
\label{momcons}
\left\langle\cos\Delta\phi\right\rangle_{\rm p_t\ cons.}=-\frac{p_{t}^{(t)}p_{t}^{(a)}}{\left\langle\sum p_t^2\right\rangle}.
\end{equation}
We assume that these are the only contributions at large relative
pseudorapidity $\Delta\eta$. In practice, we use STAR data projected
onto $|\Delta\eta|>0.7$. 


In order to isolate the flow contribution, one must quantitatively estimate the effect of momentum conservation.
We estimate the denominator of Eq.~\eqref{momcons} in the following way: 
we use the total pion and kaon multiplicities measured by BRAHMS in 
central collisions~\cite{Bearden:2004yx}, and we rescale them to the
20\%-60\% centrality range assuming that the correlation
(\ref{momcons}) scales with centrality approximately like $1/N_{\rm
 part}$, which increases the correlation by a factor $\simeq 3.6$. 
We calculate $\langle p_t^2\rangle$ for pions, kaons and nucleons 
assuming exponential $m_t$ spectra, and using the inverse slopes of
$m_t$ spectra measured by PHENIX at mid-rapidity in the centrality
range 40\%-50\%~\cite{Adler:2003cb}. 
We further estimate that the effective inverse slopes, averaged over
all rapidities, are 10\% smaller than the inverse slopes at mid-rapidity,
based on the observed decrease of $\langle p_t\rangle$ versus
rapidity~\cite{Bearden:2004yx}.
We thus obtain 
\begin{equation}
\label{momcons2}
\left\langle\cos\Delta\phi\right\rangle_{\rm p_t\ cons.}\simeq
\frac {-0.00185} {({\rm GeV}/c)^2}\, p_{t}^{(t)}p_{t}^{(a)}.
\end{equation}

Subtracting the contribution of momentum conservation from the measured
$\cos\Delta\phi$, we extract $v_{1}^{(t)} v_{1}^{(a)}$ using
Eq.~(\ref{decomposition}). Since $v_1$ is defined as a
correlation with a direction $\Phi_1$ which is itself uncorrelated
with the reaction plane~\cite{Teaney:2010vd}, one expects that this
quantity is independent of the orientation of the trigger particle.
This is indeed what is observed by analyzing STAR
results~\cite{Luzum:2010sp}. In order to increase the statistics, we
average the measured $\langle\cos\Delta\phi\rangle$ over all orientations. 
We thus obtain 10 values of $\langle\cos\Delta\phi\rangle$, corresponding to 5 intervals
in $p_t$ below 3~GeV/c for the associated particle, and 2
intervals in $p_t$ for the trigger particle, namely, 
$p_{t}^{(t)}$ = 3--4, 4--6 GeV.

\begin{figure}[]
\includegraphics[width=\linewidth]{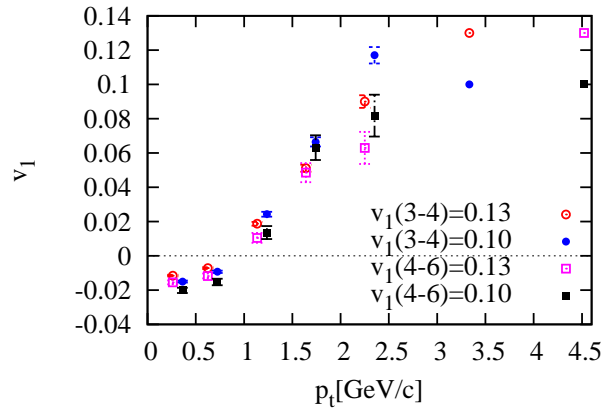}
\caption{(Color online) 
Differential $v_1$ of charged hadrons extracted from STAR correlation
data \cite{Agakishiev:2010ur}. Circles (squares) correspond to trigger particles with 
$p_t^{(t)}$ = 3--4 (4--6) GeV.
Closed (open) symbols correspond to assumed values of 
$v_1^{(t)}=0.1$ (0.13). 
Error bars are statistical only. 
The value of $p_t$ on the horizontal axis is the average value 
in the corresponding bin. For $p_t>0.5$~GeV/c, this value is
obtained using $p_t$ spectra measured by PHENIX~\cite{Adler:2003au} in
the centrality bin 30-40\%. 
For the lowest bin, it is obtained by assuming an exponential $m_t$
spectrum. 
For sake of clarity, $\langle p_t\rangle$ has been shifted by
$-0.05$ ($0.05$) for open (closed) symbols. 
}
\label{fig:v1}
\end{figure}

Assuming some value of $v_{1}^{(t)}$, we can extract $v_1^{(a)}$ from
the product $v_{1}^{(t)}v_{1}^{(a)}$. 
There is some arbitrariness in this procedure, but $v_1$ is expected
to be a smooth function of $p_t$, so that it should have comparable
values in the intervals 
2--3 and 3--4 GeV.
Any value of $v_{1}^{(t)}$ in
the range 0.10--0.13 gives reasonable results for both sets of
trigger particles, as shown in Fig.~\ref{fig:v1}. In particular, 
we obtain
similar curves for $v_1^{(a)}$ from both sets of trigger particles, which
supports the validity of Eq.~(\ref{decomposition}). 

Teaney and Yan \cite{Teaney:2010vd} used ideal hydrodynamics to make predictions for this $v_1$, and, although the values given were not intended as precise, realistic predictions, the generic  properties seen in Fig.~\ref{fig:v1} match these expectations.
The variation of $v_1$ with $p_t$ is as predicted below 2~GeV (see Fig.~12 of
Ref.~\cite{Teaney:2010vd}). 
In particular, the absence of net transverse momentum implies that the 
sum of $p_t v_1(p_t)$ over all particles vanishes, which explains why
$v_1$ changes sign near 0.8~GeV/c. 
Above 2~GeV, $v_1$ deviates from the linear rise predicted by 
ideal hydrodynamics and saturates or decreases, as expected, in the same way as 
elliptic flow~\cite{Afanasiev:2009wq}, due to viscous corrections and/or the onset of hard physics not captured by hydrodynamics. 
The magnitude of $v_1$ is also roughly as expected: the initial dipole
asymmetry $\varepsilon_1$ defined by Eq.~(\ref{defpsi1}) is predicted
to be approximately 4 times smaller than the initial eccentricity
$\varepsilon_2$ for Au-Au collisions~\cite{Teaney:2010vd}. In the
momentum range 
$1.5<p_t<2$~GeV, 
hydrodynamics predicts 
$v_1/\varepsilon_1\simeq v_2/\varepsilon_2$. 
Since $v_2\simeq 0.2$ in this momentum range for the same centrality
window~\cite{Afanasiev:2009wq}, one expects $v_1\simeq 0.05$, in good
agreement with the value in Fig.~\ref{fig:v1}. 

The estimates in Fig.~\ref{fig:v1} have sizable errors.  First, there are unnecessary statistical errors (shown in the figure),
because they are extracted from correlations involving a trigger
particle with high transverse momentum, and there are few such
particles.  Uncertainty in $\langle p_t \rangle$ for each bin is also present.  Most important, however, is the large systematic error coming from the estimation of the momentum conservation correlation.   We estimate that our value for the denominator in Eq.~\eqref{momcons} is only accurate to $\sim\pm20\%$.  This can change the extracted value of $v_1$ by as much as 50\%.

Fortunately, these errors
can be significantly reduced by carrying out a
dedicated analysis.

\section{New method of analysis}
\label{s:method}

Directed flow is most often analyzed using an event-plane method, analogous to the one often used for elliptic flow~\cite{Danielewicz:1985hn,Poskanzer:1998yz}. In every event, one defines the directed flow ``event plane\footnote{Note that $\psi_{EP,1}$, does not actually denote a plane, but is a specific direction for directed flow.  The terminology is meant to emphasize the similarity to a $v_2$ event plane analysis}'' $\psi_{EP,1}$ by 
\begin{eqnarray}
\label{eventplane}
Q\cos\psi_{EP,1}=\sum w_j \cos\phi_j\cr
Q\sin\psi_{EP,1}=\sum w_j \sin\phi_j
\end{eqnarray}
where the sum is over particles detected in the event, $\phi_j$ are the azimuthal angles of outgoing particles, and $w_j$ is a weight, and $Q\ge 0$. 
Our new method is based on a new choice for the weight $w_j$. 
The weight of a particle is generally a function of its transverse momentum $p_t$ and rapidity $y$. The desired conditions are that
\begin{itemize}
\item Correlations from momentum conservation should not bias the flow analysis. Since these correlations are proportional to $p_t$, this gives the condition $\langle w(p_t,y)p_t\rangle=0$, where angular brackets denote an average over the detector acceptance. 
\item Particles with larger $v_1$ should be given more weight, the optimal choice being $w(p_t,y)\propto v_1(p_t,y)$~\cite{Borghini:2000sa}.
\end{itemize}
For smooth initial conditions, the symmetry between target and projectile implies that $v_1$ is an odd function of $y$. A standard choice is $w(p_t,y)=y$, which satisfies both requirements if the detector acceptance is symmetric with respect to midrapidity.  By construction, this results in a measured $v_1$ that, on average, vanishes at midrapidity and is antisymmetric in $y$.

In this paper, we study directed flow created 
by fluctuations in the initial geometry. These fluctuations are expected to depend weakly on  rapidity~\cite{Bozek:2010vz}. Thus we choose a weight that depends on $p_t$ only. The choice
\begin{equation}
w(p_t)=p_t-\frac{\langle p_t^2\rangle}{\langle p_t\rangle},
\end{equation}
where angular brackets denote an average over particles in the detector acceptance, 
satisfies the condition $\langle w p_t\rangle=0$, so
that momentum conservation does not bias the analysis. 
It also corresponds to the expected behavior of $v_1(p_t)$ in ideal
hydrodynamics in the limit of low freeze-out temperature for massless
particles~\cite{Teaney:2010vd}. 


Instead of the event-plane method, one can use any of its variants such as the scalar-product method~\cite{Adler:2002pu}. All methods should give the same result provided that one only correlates particles, or subevents, separated by a gap in pseudorapidity in order to remove nonflow effects. 

\section{Conclusions}

We have shown that the first Fourier harmonic, $\langle\cos\Delta\phi\rangle$,  of the two-particle correlation measured by STAR shows the first evidence for a sizable directed flow 
originating from fluctuations of the initial geometry. 
This is an interesting new phenomenon that, like triangular flow, results from initial state fluctuations and subsequent hydrodynamic evolution, and provides the last puzzle piece needed for a complete description of long-range dihadron correlations. 
Unlike the usual directed flow, this phenomenon has no correlation with the reaction plane and should depend weakly on rapidity.  It changes sign around $p_t\simeq 0.8$~GeV, and may reach values as large as 10\%   at high $p_t$ for mid-central collisions. 

We have proposed a specific method to analyze this new observable, which eliminates correlations from momentum conservation.  
%
Such dedicated measurements of this new $v_1$ 
at RHIC and LHC will help constrain models of initial conditions and confirm the hydrodynamic behavior of the collision system. 

\begin{acknowledgments}

We thank Fuqiang Wang for providing the data, and Jean-Paul Blaizot and Frederique 
Grassi for discussions. We also thank Nicolas Borghini for noticing the typo in Eq.~\eqref{eq:dndp} in the published manuscript, which has been corrected here.
This work was funded by Agence Nationale de la Recherche under grant
ANR-08-BLAN-0093-01. 

\end{acknowledgments}



\begin{thebibliography}{99}

\bibitem{Ackermann:2000tr}
  K.~H.~Ackermann {\it et al.}  [STAR Collaboration],
  Phys.\ Rev.\ Lett.\  {\bf 86}, 402 (2001)
  [arXiv:nucl-ex/0009011].

\bibitem{Alver:2010gr}
 B.~Alver and G.~Roland,
 Phys.\ Rev.\  C {\bf 81}, 054905 (2010)
 [Erratum-ibid.\  C {\bf 82}, 039903 (2010)]
 [arXiv:1003.0194 [nucl-th]].

\bibitem{Luzum:2010sp}
  M.~Luzum,
  arXiv:1011.5773 [nucl-th].

\bibitem{Agakishiev:2010ur}
 H.~Agakishiev {\it et al.},
 arXiv:1010.0690 [nucl-ex].

\bibitem{Teaney:2010vd}
 D.~Teaney and L.~Yan,
 arXiv:1010.1876 [nucl-th].

\bibitem{Borghini:2005kd}
  N.~Borghini and J.~Y.~Ollitrault,
  Phys.\ Lett.\  B {\bf 642}, 227 (2006)
  [arXiv:nucl-th/0506045].

\bibitem{Alt:2003ab}
 C.~Alt {\it et al.}  [NA49 Collaboration],
 Phys.\ Rev.\  C {\bf 68}, 034903 (2003)
 [arXiv:nucl-ex/0303001].

\bibitem{Adams:2003zg}
 J.~Adams {\it et al.}  [STAR Collaboration],
 Phys.\ Rev.\ Lett.\  {\bf 92}, 062301 (2004)
 [arXiv:nucl-ex/0310029].

\bibitem{Back:2005pc}
 B.~B.~Back {\it et al.}  [PHOBOS Collaboration],
 Phys.\ Rev.\ Lett.\  {\bf 97}, 012301 (2006)
 [arXiv:nucl-ex/0511045].


\bibitem{Borghini:2000cm}
 N.~Borghini, P.~M.~Dinh and J.~Y.~Ollitrault,
 Phys.\ Rev.\  C {\bf 62}, 034902 (2000)
 [arXiv:nucl-th/0004026].

\bibitem{Bearden:2004yx}
 I.~G.~Bearden {\it et al.}  [BRAHMS Collaboration],
 Phys.\ Rev.\ Lett.\  {\bf 94}, 162301 (2005)
 [arXiv:nucl-ex/0403050].


\bibitem{Adler:2003cb}
 S.~S.~Adler {\it et al.}  [PHENIX Collaboration],
 Phys.\ Rev.\  C {\bf 69}, 034909 (2004)
 [arXiv:nucl-ex/0307022].

\bibitem{Adler:2003au}
 S.~S.~Adler {\it et al.}  [PHENIX Collaboration],
 Phys.\ Rev.\  C {\bf 69}, 034910 (2004)
 [arXiv:nucl-ex/0308006].


\bibitem{Afanasiev:2009wq}
 S.~Afanasiev {\it et al.}  [PHENIX Collaboration],
 Phys.\ Rev.\  C {\bf 80}, 024909 (2009)
 [arXiv:0905.1070 [nucl-ex]].



\bibitem{Danielewicz:1985hn}
 P.~Danielewicz and G.~Odyniec,
 Phys.\ Lett.\  B {\bf 157} (1985) 146.

\bibitem{Poskanzer:1998yz}
 A.~M.~Poskanzer and S.~A.~Voloshin,
 Phys.\ Rev.\  C {\bf 58}, 1671 (1998)
 [arXiv:nucl-ex/9805001].

\bibitem{Borghini:2000sa}
 N.~Borghini, P.~M.~Dinh and J.~Y.~Ollitrault,
 Phys.\ Rev.\  C {\bf 63}, 054906 (2001)
 [arXiv:nucl-th/0007063].

\bibitem{Bozek:2010vz}
 P.~Bozek, W.~Broniowski and J.~Moreira,
 arXiv:1011.3354 [nucl-th].

\bibitem{Adler:2002pu}
  C.~Adler {\it et al.}  [STAR Collaboration],
  Phys.\ Rev.\  C {\bf 66}, 034904 (2002)
  [arXiv:nucl-ex/0206001].

\end{thebibliography}
\end{document}